\documentclass[preprint,preprintnumbers, prd, floatfix, superscriptaddress,nofootinbib] {revtex4-1}
\usepackage{epsfig}
\usepackage{subfigure}
\usepackage{dcolumn}
\usepackage{bm}
\usepackage[usenames ,dvipsnames]{xcolor}
\usepackage{slashed}
\usepackage{graphicx,color}
\begin{document}
\title{Study of the $D_s^+\to a_0(980) \rho$ and $a_0(980) \omega$ decays}

\author{Yao Yu}
\email{yuyao@cqupt.edu.cn}
\affiliation{Chongqing University of Posts \& Telecommunications, Chongqing 400065, China}

\author{Yu-Kuo Hsiao}
\email{yukuohsiao@gmail.com (Corresponding author)}
\affiliation{School of Physics and Information Engineering, Shanxi Normal University, Linfen 041004, China}

\author{Bai-Cian Ke}
\email{baiciank@ihep.ac.cn}
\affiliation{School of Physics and Microelectronics,
Zhengzhou University, Zhengzhou, Henan 450001, China}

\date{\today}

\begin{abstract}
We study $D_{s}^{+}\to\rho^{0(+)}a^{+(0)}_{0}$, $D_{s}^{+}\to\omega a^{+}_{0}$,
and the resonant $D_{s}^{+}\to\rho a_0$,$a_{0}\to \eta\pi(KK)$ decays.
In the final state interaction,
where $D_s^+\to (\eta^{(\prime)}\pi^+,K^+\bar K^0)$ are followed by
the $(\eta^{(\prime)}\pi^+,K^+\bar K^0)$ to $\rho^{0(+)}a^{+(0)}_{0}$ rescatterings,
we predict ${\cal B}(D_{s}^{+}\to\rho^{0(+)}a^{+(0)}_{0})=(3.0\pm 0.3\pm 1.0)\times 10^{-3}$.
Due to the cancellation of the rescattering effects and
the suppressed short-distance $W$ annihilation contribution,
we expect that ${\cal B}(D_{s}^{+}\to\omega a^{+}_{0})
\simeq {\cal B}(D_s^+\to\pi^+\pi^0)<3.4\times 10^{-4}$.
In our calculation,
${\cal B}(D_{s}^{+}\to\rho^{0}(a^{+}_{0}\to)\eta\pi^{+})
=(1.6^{+0.2}_{-0.3}\pm 0.6)\times 10^{-3}$ agrees with the data,
whereas ${\cal B}(D_{s}^{+}\to\rho^{+}(a^{0}_{0}\to)K^+K^-)$ is
10 times smaller than the observation, which requires a careful examination.
\end{abstract}

\maketitle
\section{introduction}
The two-body $D_s^+\to PP,PV$ decays
with $P(V)$ denoting the strangeless pesudoscalar (vector) meson
have no configurations from the $W$-boson emission processes
because $\bar s$ in $D_s^+$ cannot be eliminated,
as drawn in Fig.~\ref{topy} for the topological diagrams T and C.
Interestingly, it leads to a specific exploration for
the annihilation mechanism applied to
$D_s^+\to \pi^+\pi^0,\pi^{+}\rho^{0},\pi^+\omega$~\cite{Fajfer:2003ag,
Bhattacharya:2009ps,Cheng:2010ry,Fusheng:2011tw,Cheng:2019ggx,Li:2013xsa}.

In the short-distance $W$-boson annihilation (WA)
$D_s^+(c\bar s)\to W^+\to u\bar d$ process,
$u\bar d$ can be seen to move in the opposite directions
in the $D_s^+$ rest frame, such that there exists no orbital angular momentum
between them. It indicates that $G(u\bar d)=G(\pi^+)=+1$
with $G$ denoting the $G$-parity symmetry~\cite{Cheng:2010ry}.
Since $G$-parity is a multiplicative quantum number,
one obtains $G(\pi^+\rho^0,\pi^+\pi^0)=(+1,-1)$.
Consequently, the WA $D_s^+\to \pi^+\rho^0(\pi^+\pi^0)$ decay 
due to $G(u\bar d)=G(\pi^+\rho^0)[=-G(\pi^+\pi^0)]$ 
is a $G$-parity conserved (violated) process,
which corresponds to the experimental result
${\cal B}(D_s^+\to \pi^+\rho^0)=(1.9\pm 1.2)\times 10^{-4}$
$[{\cal B}(D_s^+\to \pi^+\pi^0)<3.4\times 10^{-4}]$~\cite{pdg}.
By contrast, although the WA $D_s^+\to\pi^+\omega$ decay
violates the $G$-parity symmetry,
${\cal B}(D_s^+\to\pi^+\omega)=(1.9\pm 0.3)\times 10^{-3}$
shows no suppression~\cite{pdg}. It is hence considered to receive
the long-distance annihilation contribution~\cite{Fajfer:2003ag,Cheng:2010ry}.

The $D_s^+\to SP,SV$ decays
can help to investigate the short and long-distance
annihilation mechanisms~\cite{Cheng:2010vk,Hsiao:2019ait,Ling:2021qzl},
where $S$ stands for a non-strange scalar meson. For example,
the WA process for $D_s^+\to a_0^{0(+)}\pi^{+(0)}$
violates $G$-parity~\cite{Achasov:2017edm,Hsiao:2019ait}, such that
its branching fraction is expected as small as ${\cal B}(D_s^+\to \pi^+\pi^0)$.
Nonetheless, one measures that ${\cal B}(D_s^+\to a_0^{0(+)}\pi^{+(0)})\simeq
100 {\cal B}(D_s^+\to \pi^+\rho^0)$~\cite{BESIII:2019jjr,pdg}.
Clearly, it indicates the main contribution
from the long-distance annihilation process~\cite{Hsiao:2019ait}.
Explicitly, the long-distance annihilation process for $D_s^+\to a_0^{0(+)}\pi^{+(0)}$
starts with the $D_s^+\to\eta^{(\prime)}\rho^+$ weak decay, followed by
the $\eta^{(\prime)}$ and $\rho^+$ rescattering. With the $\pi^{+(0)}$ exchange,
$\eta^{(\prime)}$ and $\pi^+$ are turned into $a_0^{+(0)}$ and $\pi^{0(+)}$, respectively.
Since BESIII has recently reported the first observation of
the branching fractions of $D_s^+\to SV,S\to PP$
as~\cite{BESIII:2021qfo,BESIII:2021aza}
\begin{eqnarray}\label{data1}
{\cal B}_+(D_s^+\to a_0^+ \rho^0,a_0^+\to\pi^+\eta)
&=&(2.1\pm 0.8 \pm 0.5)\times 10^{-3}\,,\nonumber\\
{\cal B}_0(D_s^+\to a_0^0 \rho^+,a_0^0\to K^+ K^-)
&=&(0.7\pm 0.2\pm 0.1)\times 10^{-3}\,,
\end{eqnarray}
we are wondering which of the short and long-distance
annihilation processes can be the dominant contribution.
Hence, we propose to study
$D_s^+\to a_0^{+(0)} \rho^{0(+)}$, $D_s^+\to a_0^+\omega$,
and the resonant three-body $D_s^+\to a_0^{+(0)} \rho^{0(+)},
a_0^{+(0)}\to[\eta\pi^{+(0)},K^+\bar K^0(K^+ K^-)]$ decays,
in order to analyze the data in Eq.~(\ref{data1}). We will also test if
${\cal B}(D_s^+\to a_0^+ \rho^0, a_0^+ \rho^0)$ have nearly equal sizes as
${\cal B}(D_s^+\to a_0^+ \pi^0)\simeq {\cal B}(D_s^+\to a_0^0 \pi^+)$
that respects the isospin symmetry.

\section{Formalism}
%
\begin{figure}[t!]
\includegraphics[width=1.8in]{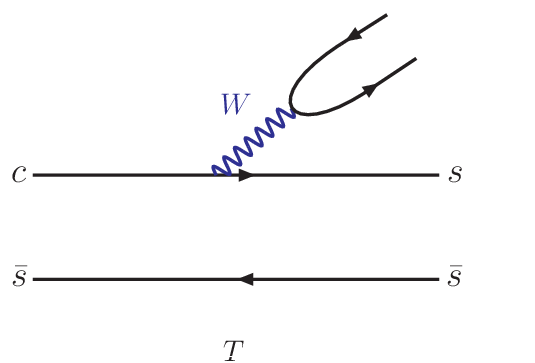}
\includegraphics[width=1.8in]{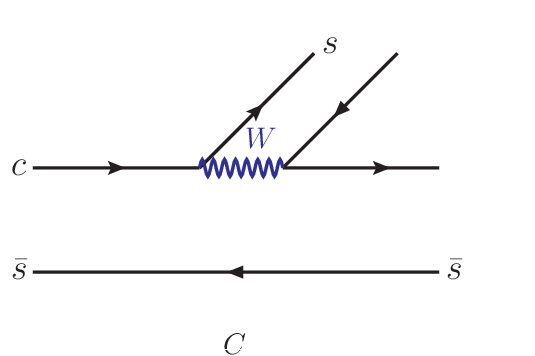}
\includegraphics[width=1.8in]{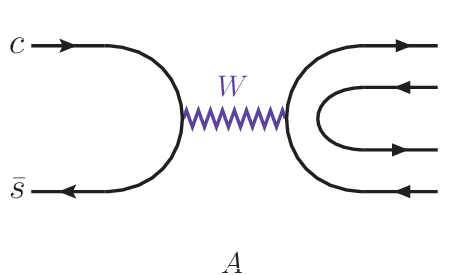}
\caption{Topological diagrams for the Cabibbo-allowed $D_s^+$ weak decays.}\label{topy}
\end{figure}
%
\begin{figure}[t!]
\includegraphics[width=2.1in]{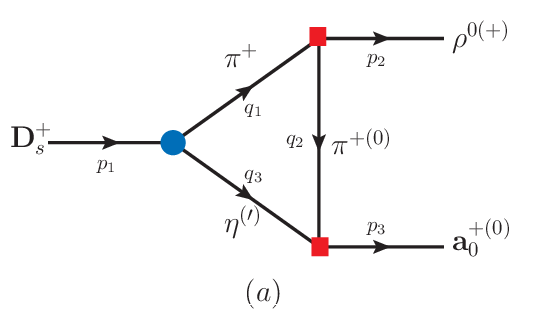}\\
\includegraphics[width=2.1in]{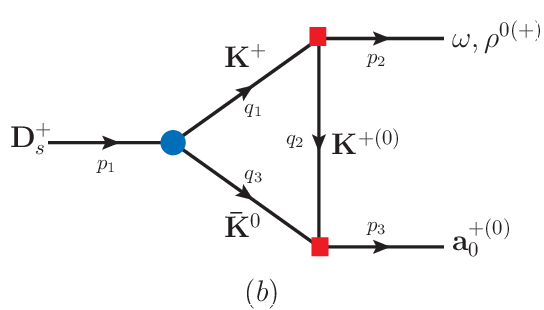}
\includegraphics[width=2.1in]{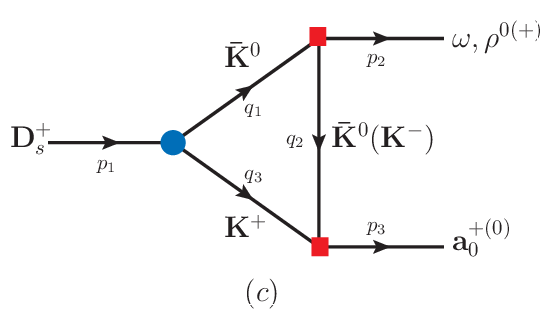}
\caption{Triangle rescattering diagrams
for $D^+_s\to\rho^{0(+)}a_0^{+(0)}$ and $D^+_s\to\omega a_0^+$.}\label{triangle}
\end{figure}
%
Considering the short-distance WA processes,
$D_s^+\to PP(PV)$ and $D_s^+\to SP(SV)$
both get an $A$ term as the annihilation amplitude in Fig.~\ref{topy}.
According to ${\cal B}(D_s^+\to \pi^+\rho^0)\simeq 10^{-4}$
that receives the short-distance WA contribution~\cite{pdg},
one regards $A$ to give ${\cal B}(D_s^+\to a_0^{+(0)} \rho^{0(+)})$
not larger than $10^{-4}$. However, ${\cal B}_+$ in Eq.~(\ref{data1})
suggests ${\cal B}(D_s^+\to a_0^+ \rho^0)\simeq 10^{-3}$.
This strongly suggests that the main contribution
to $D_s^+\to a_0^{+(0)} \rho^{0(+)}$ is from the long-distance annihilation process.
For $D_s^+\to a_0^+\omega$,
the WA  contribution is suppressed with the $G$-parity violation,
such that $A\simeq 0$. Therefore, we start with the triangle rescattering processes
for $D_s^+\to a_0^{+(0)} \rho^{0(+)}$ and $D_s^+\to a_0^+\omega$
 as the most possible main contributions.

See Fig.~\ref{triangle}, the rescattering processes for $D_s^+\to a_0^{+(0)} \rho^{0(+)}$
include both the weak and strong decays.
In our case, the weak decays come from
$D^{+}_{s}\to \pi^{+}\eta^{(\prime)},K^{+}\bar{K}^0$,
and the amplitudes are given
by~\cite{Cheng:2010ry,Cheng:2019ggx,Fusheng:2011tw,Bhattacharya:2009ps}
\begin{eqnarray}\label{M1}
{\cal M}_\eta(D^{+}_{s}\to \pi^{+}\eta)
&=&
\frac{G_F}{\sqrt{2}} V^*_{cs}V_{ud}(\sqrt{2}A\cos\phi-T\sin\phi)\,,\nonumber\\
{\cal M}_{\eta'}(D^{+}_{s}\to \pi^{+}\eta^{\prime})
&=&
 \frac{G_F}{\sqrt{2}}V^*_{cs}V_{ud}(\sqrt{2}A\sin\phi+T\cos\phi)\,,\nonumber\\
{\cal M}_K(D^{+}_{s}\to K^{+}\bar{K}^0)
&=&
\frac{G_F}{\sqrt{2}}V^*_{cs}V_{ud}(C+A)\,,
\end{eqnarray}
where $G_F$ is the Fermi constant,
$V^*_{cs}V_{ud}$ the Cabibbo-Kobayashi-Maskawa (CKM) matrix elements,
and $V^*_{cs}V_{ud}\simeq 1$ presents the Cabibbo-allowed decay modes.
In addition, $(T,C,A)$ are the topological amplitudes,
along with the mixing angle $\phi=43.5^\circ$ 
from the $\eta-\eta^\prime$ mixing matrix~\cite{FKS,FKS2}:
\begin{eqnarray}\label{eta_mixing}
\left(\begin{array}{c} \eta \\ \eta^\prime \end{array}\right)=
\left(\begin{array}{cc} \cos\phi & -\sin\phi \\ \sin\phi & \cos\phi \end{array}\right)
\left(\begin{array}{c} \sqrt{1/2}(u\bar u+d\bar d) \\ s\bar s \end{array}\right)\,.
\end{eqnarray}
For the strong decays $V,S\to PP$, the amplitudes are given by~\cite{Hsiao:2019ait}
\begin{eqnarray}\label{strong}
&&{\cal M}_{\rho^{0(+)}\to\pi^{+(0)}\pi^{-(+)}}=g_\rho\epsilon \cdot  (q_1-q_2),\,\nonumber\\
&&{\cal M}_{a_0^+\to \eta^{(\prime)}\pi^+,K^+\bar K^0}=g_{\eta^{(\prime)}},g_K\,,
\end{eqnarray}
where $\epsilon_{\mu}$ is the polarization four vector of the $\rho$ meson,
and $q_{1,2}^\mu$ are the four momenta of $\pi^+\pi^-(\pi^0\pi^+)$, respectively.
The $SU(3)$ flavor symmetry is able to relate
different $V(S)\to PP$ decay channels~\cite{Tornqvist:1979hx,Fayyazuddin:2012qfa},
such that we obtain
$g_\rho/2$ for ${\cal M}_{\rho^0\to K^+ K^-,\bar K^0 K^0}$,
$ g_\rho/\sqrt 2$ for ${\cal M}_{\rho^+\to K^+\bar K^0}$, and
$(-)g_K/\sqrt 2 $ for ${\cal M}_{a^0_0\to K^+ K^-(\bar K^0 K^0)}$,
together with $(-)g_\rho/2$ for ${\cal M}_{\omega\to K^+ K^-(\bar K^0 K^0)}$.

By assembling the weak and strong couplings in the rescattering processes,
we derive that
\begin{eqnarray}\label{m4}
{\cal M}(D_{s}^{+}\to\rho^{0(+)}a^{+(0)}_{0})
&=&{\cal M}_a+{\cal M}_{a}^{\prime}+{\cal M}_{b}+{\cal M}_{c}\,,
\nonumber\\
{\cal M}(D_{s}^{+}\to\omega a^{+}_{0})
&=& \hat {\cal M}_{b}+ \hat {\cal M}_{c}\,,
\end{eqnarray}
with $M_a^{(\prime)}$ and $M_{b(c)}[\hat M_{b(c)}]$
for Fig.~\ref{triangle}a and Fig.~\ref{triangle}b(c),
respectively. More explicitly, ${\cal M}_{a,b,c}$
are given by~\cite{Hsiao:2019ait,Yu:2020vlt,Hsiao:2021tyq}
\begin{eqnarray}\label{m5}
{\cal M}_{a}&=&\int \frac{d^4{q}_{1}}{(2\pi)^{4}}
\frac{{\cal M}_{\eta}{\cal M}_{\rho^{0(+)}\to\pi^+\pi^{-(0)}}
{\cal M}_{a^{+(0)}_{0}\to\eta\pi^{+(0)}} F_{\pi}(q_{2}^2)}
{(q_{1}^{2}-m_{\pi}^{2}+i\epsilon)[(q_1-p_2)^{2}-m_{\pi}^{2}+i\epsilon][(q_1-p_1)^{2}-m_{\eta}^{2}+i\epsilon]}\,,\nonumber\\
{\cal M}_{b}&=&\int \frac{d^4{q}_{1}}{(2\pi)^{4}}
\frac{{\cal M}_{K} {\cal M}_{\rho^{0(+)}\to K^+K^-(K^+\bar K^0)}
{\cal M}_{a_0^{+(0)}\to \bar{K}^0 K^{+(0)}} F_{K}(q_{2}^2)}
{(q_{1}^{2}-m_{K}^{2}+i\epsilon)[(q_1-p_2)^{2}-m_{K}^{2}+i\epsilon][(q_1-p_1)^{2}-m_{K}^{2}+i\epsilon]}\,,
\nonumber\\
{\cal M}_{c}&=&\int \frac{d^4{q}_{1}}{(2\pi)^{4}}
\frac{{\cal M}_K {\cal M}_{\rho^{0(+)}\to  \bar{K}^0 K^{0(+)}}
{\cal M}_{a_0^{+(0)}\to \bar K^0 K^{+(0)}} F_{K}(q_{2}^2)}
{(q_{1}^{2}-m_{K}^{2}+i\epsilon)[(q_1-p_2)^{2}-m_{K}^{2}+i\epsilon][(q_1-p_1)^{2}-m_{K}^{2}+i\epsilon]}\,,
\end{eqnarray}
with $q_2=q_1-p_2$ and $q_3=q_1-p_1$ following the momentum flows in Fig.~\ref{triangle}.
The form factor
$F_M(q_{2}^2)\equiv(\Lambda_M^{2}-m^{2}_M)/(\Lambda_M^{2}-q^{2}_{2})$
with the cutoff parameter $\Lambda_M$ [$M=(\pi,K)$] is to avoid
the overestimation with $q_2$ to $\pm\infty$~\cite{Du:2021zdg}.
Substituting $\eta^\prime$ and $\omega$
for $\eta$ in ${\cal M}_a$ and $\rho^0$ in ${\cal M}_{b(c)}$
leads to ${\cal M}_a^\prime$ and $\hat {\cal M}_{b(c)}$, respectively.

As a consequence, we find that
\begin{eqnarray}\label{m6}
{\cal M}(D_{s}^{+}\to\rho^{0}a^{+}_{0})&=&{\cal M}(D_{s}^{+}\to\rho^{+}a^{0}_{0})\,,\nonumber\\
{\cal M}(D_{s}^{+}\to\omega a^{+}_{0})&=&0\,,
\end{eqnarray}
where the first relation respects the isospin symmetry,
whereas $\hat {\cal M}_{b}=- \hat {\cal M}_{c}$ due to
${\cal M}_{\omega\to K^+ K^-}=-{\cal M}_{\omega\to \bar K^0 K^0}$
cancels the rescattering contributions to $D_{s}^{+}\to\omega a^{+}_{0}$,
which causes ${\cal M}(D_{s}^{+}\to\omega a^{+}_{0})=0$.

To deal with the triangle loops in Eq.~(\ref{m5}),
the equation in Refs.~\cite{tHooft:1978jhc,Hahn:1998yk,Denner:2005nn,Passarino:1978jh} can be useful,
given by
\begin{eqnarray}\label{int}
&&
\int \frac{d^{4}q_1}{i \pi^2}\frac{q_1^{\mu}}
{(q_1^{2}-m_{1}^{2}+i\epsilon)[(q_1-p_2)^{2}-m_{2}^{2}+i\epsilon][(q_1-p_1)^{2}-m_{3}^{2}+i\epsilon]}\nonumber\\
&&=
-p_\rho^\mu C_1(p_{a_0}^2,m_\rho^2,m_{D_s}^2,m_1^2,m_2^2,m_3^2)
-p_{D_s}^\mu C_2(p_{a_0}^2,m_\rho^2,m_{D_s}^2,m_1^2,m_2^2,m_3^2)\,.
\end{eqnarray}
By calculating the triangle rescattering processes,
we obtain
\begin{eqnarray}
&&
\Gamma(D_{s}^{+}\to\rho^{0{+}}a^{+(0)}_{0})=
\frac{|\vec{p}_{2}|^{3}}{8\pi m_{\rho}^2}|H|^2\,,\nonumber\\
&&
H=\frac{1}{8\pi^2}\bigg(g_\rho g_\eta{\cal M}_\eta C_\eta+g_\rho g_{\eta'}{\cal M}_{\eta'} C_{\eta'}
+g_\rho g_K {\cal M}_K C_K\bigg)\,,
\end{eqnarray}
with $C_{\eta^{(\prime)}}=C_{2,\eta^{(\prime)}}-C_{2,\eta^{(\prime)}}^*$ and
$C_K=C_{2,K}-C_{2,K}^*$ as the integrated results of the $C_2$ terms in Eq.~(\ref{int}),
where $C_{2,M}^{(*)}$ are defined by
\begin{eqnarray}
C_{2,\eta}^{(*)}&=&C_2(m_\pi,m_\pi(\Lambda_\pi),m_\eta)\,,\nonumber\\
C_{2,\eta'}^{(*)}&=&C_2(m_\pi,m_\pi(\Lambda_\pi),m_{\eta'})\,,\nonumber\\
C_{2,K}^{(*)}&=&C_2(m_K,m_K(\Lambda_K),m_K)\,.
\end{eqnarray}
However, the $C_1$ terms have been disappearing
due to $p_\rho\cdot \epsilon=0$ in the amplitudes.

For the three-body decays
$D_{s}^{+}\to\rho^{0(+)}(a^{+(0)}_{0}\to) \eta\pi^{+(0)},K^+\bar K^0(K^+ K^-)$,
we present~\cite{pdg}
\begin{eqnarray}\label{gamma1}
&&\Gamma(D_{s}^{+}\to\rho a_0,a_0\to\eta\pi(KK))\nonumber\\
&&=\int\int
\frac{1}{(2\pi)^3}\frac{|\vec{p}_{2}|^{2}}{32m_{D_{s}} m_{\rho}^2}|H|^2
\frac{g_{\eta(K)}^{2}}{\mid D_{a_{0}}(s)\mid^{2}}ds dt\,,
\end{eqnarray}
with $s=(p_{\eta(K)}+p_{\pi(K)})^{2}$ and $t=(p_{\rho}+p_{\pi(K)})^{2}$,
such that the theoretical results can be compared to the data in Eq.~(\ref{data1}).
In the above equation, $1/D_{a_{0}}(s)$ presents the propagator for $a_0$,
and we define~\cite{Achasov:2004uq}
\begin{eqnarray}\label{Da01}
&&D_{a_{0}}(s)=s-m_{a_0}^{2}-\sum\limits_{\alpha\beta}
{\bigg[\text{Re}\Pi^{\alpha\beta}_{a_0}(m^2_{a_0})-\Pi^{\alpha\beta}_{a_0}(s)\bigg]}\,,
\end{eqnarray}
where
\begin{eqnarray}\label{Da02}
\Pi^{\alpha\beta}_{a_0}(x)&=&
\frac{G^{2}_{\alpha\beta}}{16\pi}
\bigg\{\frac{m_{\alpha\beta}^+ m_{\alpha\beta}^-}{\pi x}
\log\bigg[\frac{m_\beta}{m_\alpha}\bigg]-\theta[x-(m_{\alpha\beta}^+)^2]\nonumber\\
&\times&
\rho_{\alpha\beta}\bigg(i+\frac{1}{\pi}
\log\bigg[\frac{\sqrt{x-(m_{\alpha\beta}^+)^{2}}+\sqrt{x-(m_{\alpha\beta}^-)^{2}}}
{\sqrt{x-(m_{\alpha\beta}^-)^{2}}-\sqrt{x-(m_{\alpha\beta}^+)^{2}}}\bigg]\bigg)
\nonumber\\
&-&\rho_{\alpha\beta}\bigg(1-\frac{2}{\pi}
\arctan\bigg[\frac{\sqrt{-x+(m_{\alpha\beta}^+)^{2}}}{\sqrt{x-(m_{\alpha\beta}^-)^{2}}}\bigg]\bigg)
(\theta[x-(m_{\alpha\beta}^-)^{2}]-\theta[x-(m_{\alpha\beta}^+)^{2}])\nonumber\\
&+&\rho_{\alpha\beta}\frac{1}{\pi}
\log\bigg[\frac{\sqrt{(m_{\alpha\beta}^+)^{2}-x}+\sqrt{(m_{\alpha\beta}^-)^2-x}}
{\sqrt{(m_{\alpha\beta}^-)^{2}-x}-\sqrt{(m_{\alpha\beta}^+)^2-x}}\bigg]
\theta[(m_{\alpha\beta}^-)^{2}-x]\bigg\}\,,\nonumber\\
\rho_{\alpha\beta}&\equiv& \left|\sqrt{x-(m_{\alpha\beta}^+)^{2}}\sqrt{x-(m_{\alpha\beta}^-)^{2}}\right|/x\,,
\end{eqnarray}
with $G_{\alpha\beta}=(g_{\eta^{(\prime)}},g_K)$
for $\alpha\beta={(\eta^{(\prime)}\pi,K \bar K)}$ and $m_{\alpha\beta}^\pm=m_\alpha\pm m_\beta$.

\section{Numerical Results}
In the numerical analysis, we adopt
$V_{cs}=V_{ud}=1-\lambda^2/2$ with $\lambda=0.22453\pm 0.00044$
in the Wolfenstein parameterization~\cite{pdg},
along with $m_{a_0^0}=0.987$~GeV~\cite{Kornicer:2016axs,Bugg:2008ig}.
The topological parameters $(T,C,A)$ have been extracted as~\cite{Cheng:2019ggx}
\begin{eqnarray}\label{TCA}
&&
(|T|, |C|, |A|)=(0.363\pm0.001,0.323\pm0.030,0.064\pm0.004)~\mbox{GeV}^{3}~\,,\nonumber\\
&&
(\delta_C,\delta_A)=(-151.3\pm0.3,23.0^{+\;\;7.0}_{-10.0})^{\circ}\,,
\end{eqnarray}
where $\delta_{C,A}$ are the relative strong phases.
According to the extraction, we obtain
${\cal B}(D_s^+\to \pi^{+}\eta,\pi^+\eta^{\prime},K^{+}\bar{K}^0)=
(1.8\pm0.2,4.2\pm0.5,3.1\pm0.5)\times 10^{-2}$,
consistent with the experimental values of
$(1.68\pm0.10,3.94\pm0.25,2.95\pm0.14)\times10^{-2}$, respectively~\cite{pdg}.
For $V,S\to PP$, the strong coupling constants
read~\cite{Bugg:2008ig,Kornicer:2016axs,pdg}
\begin{eqnarray}
&&g_{\rho}=6.0\,,
(g_\eta,g_{\eta'},g_K)=(2.87\pm 0.09,-2.52\pm 0.08,2.94\pm 0.13)~\text{GeV}\,.
\end{eqnarray}
Empirically, $\Lambda_M$ of ${\cal O}(1.0~\text{GeV})$
is commonly used to explain the data~\cite{Tornqvist:1993ng,Li:1996yn,Wu:2019vbk};
besides, it is obtained that $\Lambda_K-\Lambda_\pi=m_K-m_\pi$~\cite{Cheng:2004ru}.
Therefore, we are allowed to use
$(\Lambda_\pi,\Lambda_K)=(1.25\pm0.25,1.60\pm 0.25)$~GeV,
which result in
\begin{eqnarray}\label{C_M}
C_{\eta} &=&[(0.57\pm 0.08)+i(0.17\pm 0.09)]~\text{GeV}^{-2}\,,\nonumber\\
C_{\eta^{\prime}} &=&[(0.34\pm 0.03)-i(0.27\pm 0.07)]~\text{GeV}^{-2}\,,\nonumber\\
C_{K} &=&[(0.85\pm 0.03)-i(0.45\pm 0.08)]~\text{GeV}^{-2}\,,
\end{eqnarray}
with $p_{a_0}^2=m_{a_0}^2$. Subsequently,
we predict
\begin{eqnarray}\label{Aab7}
&&{\cal B}(D_{s}^{+}\to\rho^{0(+)}a^{+(0)}_{0})
=(3.0\pm 0.3\pm 1.0)\times 10^{-3}\,,\nonumber\\
&&{\cal B}(D_{s}^{+}\to\omega a^{+}_{0})=0\,,
\end{eqnarray}
where the first error in ${\cal B}(D_{s}^{+}\to\rho^{0(+)}a^{+(0)}_{0})$
takes into account the uncertainties from $\Lambda_{\pi}$ and $\Lambda_{K}$,
and the second one combines those from
$V_{cs}^*$, $V_{ud}$, $(T,C,A)$, and the strong coupling constants.
For the resonant three-body decays, we obtain
\begin{eqnarray}\label{3b_result}
{\cal B}(D_{s}^{+}\to\rho^{0(+)}(a^{+(0)}_{0}\to)\eta\pi^{+(0)})
&=& (1.6^{+0.2}_{-0.3}\pm 0.6)\times 10^{-3}\,,\nonumber\\
{\cal B}(D_{s}^{+}\to\rho^{+}(a^{0}_{0}\to)K^+K^-,K^0\bar K^0)
&=& (0.9^{+0.1}_{-0.1}\pm 0.4,0.7^{+0.1}_{-0.1}\pm 0.3)\times 10^{-4}\,,\nonumber\\
{\cal B}(D_{s}^{+}\to\rho^{0}(a^{+}_{0}\to)K^+\bar{K}^0)
&=&(1.5^{+0.2}_{-0.3}\pm 0.6)\times 10^{-4}\,,
\end{eqnarray}
where the sources of the two errors are the same as those in Eq.~(\ref{Aab7}).

\section{Discussions and Conclusions}
In the triangle loop, when
the momentum flow approaches the mass shell for one of the three propagators,
the integration with $i\epsilon\simeq im\Gamma$ gives rise to the imaginary parts in Eq.~(\ref{C_M}),
where $\Gamma$ is a very tiny decay width for $\pi$, $\eta^{(\prime)}$ or $K$.
The off-shell integrations are responsible for the real parts in Eq.~(\ref{C_M}),
for which we take ${\cal M}_a$ as our description.
In principle, the integration allows a momentum flow from $-\infty$ to $+\infty$.
However, when the exchange particle proceeds with $q_2^2$ around several~GeV$^2$,
instead of the infinity, the integration with $q_2^2\sim\pm\infty$
causes an overestimation~\cite{Du:2021zdg}.
We have accordingly introduced the form factor $F_\pi$ in Eq.~(\ref{m5})
to cut off the contribution from $q_2^2\sim\pm\infty$.
While $q_1^2$, $q_2^2$ and $q_3^2$ have been associated in the loop,
$F_\pi$ also works to cut off the contributions from the propagators of
 the rescattering particles $\pi^+$ and $\eta^{(\prime)}$.
Note that the single cutoff form factor has also been commonly used
elsewhere~\cite{Wu:2021udi,Wu:2019rog,
Li:2014pfa}~\footnote{Please also consult Refs.~\cite{Cheng:2021nal},
where one considers three cutoff form factors.}.

The smallness of the WA $D_s^+\to MM$ decay can be traced back to
its amplitude, given by~\cite{Hsiao:2019wyd,Hsiao:2014zza,Huang:2021qld}

\begin{eqnarray}
&&
{\cal M}_{WA}(D_s^+\to MM)
\propto f_{D_s} q^\mu\langle MM|\bar u\gamma_\mu(1-\gamma_5) d|0\rangle\nonumber\\
&&\simeq(m_u+m_d)\langle MM|\bar u\gamma_5 d|0\rangle\,,
\end{eqnarray}
where $q^\mu\langle MM|u\gamma_\mu d|0\rangle=0$
corresponds to the conservation of the vector current (CVC);
most importantly, $m_{u(d)}\simeq 0$ causes the chiral suppression
of the WA $D_s^+\to MM$ decay. According to the data,
${\cal B}(D_s^+\to \pi^+\rho^0)=(1.9\pm 1.2)\times 10^{-4}$
indicates that ${\cal B}_{WA}(D_s^+\to MM)$ should be around $10^{-4}$.
In addition, ${\cal B}(D_s^+\to \pi^+\pi^0)<3.4\times 10^{-4}$
suggests that the G-parity violation suppresses the WA process even more.
Therefore, since $D_s^+\to \rho a_0,\rho\pi$ are both the G-parity conserved processes,
it is reasonable to present that ${\cal B}_{WA}(D_s^+\to \rho a_0)
\simeq {\cal B}_{WA}(D_s^+\to\rho \pi)\sim 10^{-4}$.
As a theoretical support,
we present
\begin{eqnarray}
&&{\cal B}_{WA}(D_s^+(c\bar s)\to\rho a_0)\nonumber\\
&=&R_f\frac
{(V_{cs}^* V_{ud}f_{D_s})^2\tau_{D_s}}
{(V_{cb}V_{ud}^* f_{B_c})^2\tau_{B_c}}
{\cal B}_{WA}(B_c^+(c\bar b)\to\rho a_0)
\sim 10^{-4}\,,
\end{eqnarray}
in agreement with our estimation, where
$R_f=3.1\times 10^{-2}$ is mostly from the phase space factors,
and ${\cal B}_{WA}(B_c^+\to\rho a_0)\simeq 1.0\times 10^{-5}$
is adopted from Ref.~\cite{Liu:2010kq}.

Disregarding the WA contributions, we predict
${\cal B}(D_{s}^{+}\to\rho^{0(+)}a^{+(0)}_{0})=3.0\times 10^{-3}$ in Eq.~(\ref{Aab7}).
It is found that the $\eta\pi,\eta'\pi,K^+\bar K^0$ rescatterings
and their interferences
give 6\%, 7\%, 30\% and 59\% of ${\cal B}(D_s^+\to a_0^{+(0)}\rho^{0(+)})$,
respectively. By contrast, the $\rho\eta^{(\prime)}$ rescatterings
from $D_s^+\to \rho\eta^{(\prime)}$
dominantly contribute to $D_s^+\to a_0^{+(0)}\pi^{0(+)}$,
instead of $D_s^+\to K^* K$~\cite{Hsiao:2019ait}.
Since $D_{s}^{+}\to\omega a^{+}_{0}$ has no rescattering effects; besides,
the WA contribution is suppressed by the $G$-parity violation,
we anticipate that ${\cal B}(D_{s}^{+}\to\omega a^{+}_{0})
\simeq {\cal B}(D_s^+\to\pi^+\pi^0)<3.4\times 10^{-4}$~\cite{pdg}.

In Eq.~(\ref{3b_result}), ${\cal B}(D_{s}^{+}\to\rho^{0}(a^{+}_{0}\to)\eta\pi^{+})
=(1.6^{+0.2}_{-0.3}\pm 0.6)\times 10^{-3}$ is able to explain the data
[see Eq.~(\ref{data1})],
demonstrating the sufficient long-distance annihilation contribution.
It is confusing that ${\cal B}(D_s^+\to a_0^0 \rho^+,a_0^0\to K^+ K^-)$
is 10 times smaller than ${\cal B}_0$ in Eq.~(\ref{data1}).
For clarification, we take the approximate form of the resonant branching fraction:
${\cal B}(D_s^+\to SV,S\to PP)\simeq {\cal B}(D_s^+\to SV){\cal B}(S\to PP)$,
together with the isospin relation:
${\cal B}(D_{s}^{+}\to\rho^0 a^+_0)\simeq {\cal B}(D_{s}^{+}\to\rho^+ a^0_{0})$,
such that ${\cal B}_0/{\cal B}_+$ is reduced as
${\cal B}(a_0^0\to K^+ K^-)/{\cal B}(a_0^+\to \eta\pi^+)\simeq 1/3$,
disagreeing with
${\cal B}(a_0^0\to K^+ K^-)/{\cal B}(a_0^+\to \eta\pi^+)=1/10$
from the experimental extraction~\cite{Cheng:2010vk}.
Therefore, we conclude that
there exists a possible contradiction between the observations in Eq.~(\ref{data1}).

In our reasoning, the contradiction might be caused by
$D_s^+\to \rho^+ f_0,f_0\to K^+ K^-$ with $f_0\equiv f_0(980)$,
which can be mistaken as $D_s^+\to \rho^+ a_0^0,a_0^0\to K^+ K^-$~\cite{PC}.
First, since $D_s^+\to \rho^+ f_0$ is an external $W$-boson emission process,
its branching fraction can be of order~$10^{-3}$. Second,
$a_0$ and $f_0$ are both scalar mesons,
and have nearly the same masses and overlapped decay widths.
As a result, it is possible that one cannot distinguish between
the resonant signals of $D_s^+\to \rho^+(f_0,a_0), (f_0,a_0)\to K^+ K^-$
in the $K^+ K^-$ invariant mass spectrum.
For a careful examination, we suggest a measurement of
${\cal B}(D_{s}^{+}\to\rho^{0}(a^{+}_{0}\to)K^+\bar{K}^0)/
{\cal B}(D_{s}^{+}\to\rho^{+}(a^{0}_{0}\to)K^+ K^-)$,
which will be observed around 2
if there exists no resonant $f_0\to K^+ K^-$ decay
to be involved in ${\cal B}(D_{s}^{+}\to\rho^+ K^+ K^-)$.

In summary, we have studied $D_{s}^{+}\to\rho^{0(+)}a^{+(0)}_{0}$,
$D_{s}^{+}\to \omega a_0^0$, and the resonant $D_{s}^{+}\to\rho a_0$,
$a_{0}\to \eta\pi(KK)$ decays. In the final state interaction,
where $D_s^+\to \eta^{(\prime)}\pi^+ (K^+\bar K^0)$ is followed by
the $\eta^{(\prime)}\pi^+$ ($K^+\bar K^0$) to $\rho a_0$ rescattering,
we have predicted
${\cal B}(D_{s}^{+}\to\rho^{0(+)}a^{+(0)}_{0})=(3.0\pm 0.3\pm 1.0)\times 10^{-3}$.
Because of the non-contribution of the rescattering effects and
the suppressed short-distance $W$ annihilation, it has been expected that
${\cal B}(D_{s}^{+}\to\omega a^{+}_{0})
\simeq {\cal B}(D_s^+\to\pi^+\pi^0)<3.4\times 10^{-4}$.
For the resonant three-body decay,
${\cal B}(D_{s}^{+}\to\rho^{0}(a^{+}_{0}\to)\eta\pi^{+})
=(1.6^{+0.2}_{-0.3}\pm 0.6)\times 10^{-3}$ has been shown to agree with the data.
We have presented ${\cal B}(D_{s}^{+}\to\rho^{+}(a^{0}_{0}\to)K^+K^-)$
10 times smaller than the observation,
indicating that a more careful examination is needed.

\section*{ACKNOWLEDGMENTS}
We would like to thank Prof.~Liang Sun for useful discussions.
YKH was supported in part by NSFC (Grant Nos.~11675030 and 12175128).
YY was supported in part by NSFC (Grant Nos.~11905023 and 12047564),
the Fundamental Research Funds for the Central Universities (Grant No.~2020CDJQY-Z003) 
and CQCSTC (Grant Nos.~cstc2020jcyj-msxmX0555 and cstc2020jcyj-msxmX0810).
BCK was supported in part by NSFC (Grant No.~11875054) and
the Chinese Academy of Sciences (CAS) Large-scale Scientific Facility Program;
Joint Large-Scale Scientific Facility Fund of the NSFC and CAS (Contract
No.~U2032104).

\end{document}